\documentclass[twoside,twocolumn,9pt]{article}
\usepackage{extsizes}
\usepackage[super,sort&compress,comma]{natbib} 
\usepackage[version=3]{mhchem}
\usepackage[left=1.5cm, right=1.5cm, top=1.785cm, bottom=2.0cm]{geometry}
\usepackage{balance}
\usepackage{mathptmx}
\usepackage{sectsty}
\usepackage{graphicx}
\usepackage{lipsum} 
\usepackage{lastpage}
\usepackage{xcolor}
\usepackage{bm}
\usepackage{color}
\usepackage[format=plain,justification=justified,singlelinecheck=false,font={stretch=1.125,small,sf},labelfont=bf,labelsep=space]{caption}
\usepackage{float}
\usepackage{fancyhdr}
\usepackage{fnpos}
\usepackage[english]{babel}
\addto{\captionsenglish}{%
  
}
\usepackage{array}
\usepackage{droidsans}
\usepackage{charter}
\usepackage[T1]{fontenc}
\usepackage[usenames,dvipsnames]{xcolor}
\usepackage{setspace}
\usepackage[compact]{titlesec}
\usepackage{hyperref}
\usepackage{ulem} 
\definecolor{cream}{RGB}{222,217,201}

\begin{document}

\pagestyle{fancy}
\thispagestyle{plain}
\fancypagestyle{plain}{
\renewcommand{\headrulewidth}{0pt}
}

\makeFNbottom
\makeatletter
\renewcommand\LARGE{\@setfontsize\LARGE{15pt}{17}}
\renewcommand\Large{\@setfontsize\Large{12pt}{14}}
\renewcommand\large{\@setfontsize\large{10pt}{12}}
\renewcommand\footnotesize{\@setfontsize\footnotesize{7pt}{10}}
\makeatother

\renewcommand{\thefootnote}{\fnsymbol{footnote}}
\renewcommand\footnoterule{\vspace*{1pt}%
\color{cream}\hrule width 3.5in height 0.4pt \color{black}\vspace*{5pt}} 
\setcounter{secnumdepth}{5}

\makeatletter 
\renewcommand\@biblabel[1]{#1}            
\renewcommand\@makefntext[1]%
{\noindent\makebox[0pt][r]{\@thefnmark\,}#1}
\makeatother 
\renewcommand{\figurename}{\small{Fig.}~}
\sectionfont{\sffamily\Large}
\subsectionfont{\normalsize}
\subsubsectionfont{\bf}
\setstretch{1.125} 
\setlength{\skip\footins}{0.8cm}
\setlength{\footnotesep}{0.25cm}
\setlength{\jot}{10pt}
\titlespacing*{\section}{0pt}{4pt}{4pt}
\titlespacing*{\subsection}{0pt}{15pt}{1pt}

\newcommand{\vecf}[2]{\bm{#1}(\bm{#2})} 
\newcommand{\ie}[1]{ #1^{(i)}} 
\newcommand{\ar}[1]{ #1^{(e)}} 
\newcommand{\dd}[2]{\frac{\partial #1}{\partial #2}}
\newcommand{\pvec}[1]{\bm{#1}'} 
\newcommand{\red}[1]{{\color{red}#1}}
\newcommand{\izs}[1]{\red{\sout{#1}}}
\newcommand{\andris}{\textcolor{magenta}}
\newcommand{\vect}[1]{\boldsymbol{#1}} 
\newcommand{\matr}[1]{\mathrm{\boldsymbol{#1}}} 

\fancyfoot{}
\fancyfoot[LO,RE]{\vspace{-7.1pt}\includegraphics[height=9pt]{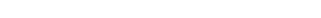}}
\fancyfoot[CO]{\vspace{-7.1pt}\hspace{13.2cm}\includegraphics{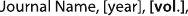}}
\fancyfoot[CE]{\vspace{-7.2pt}\hspace{-14.2cm}\includegraphics{RF}}
\fancyfoot[RO]{\footnotesize{\sffamily{1--\pageref{LastPage} ~\textbar  \hspace{2pt}\thepage}}}
\fancyfoot[LE]{\footnotesize{\sffamily{\thepage~\textbar\hspace{3.45cm} 1--\pageref{LastPage}}}}
\fancyhead{}
\renewcommand{\headrulewidth}{0pt} 
\renewcommand{\footrulewidth}{0pt}
\setlength{\arrayrulewidth}{1pt}
\setlength{\columnsep}{6.5mm}
\setlength\bibsep{1pt}

\makeatletter 
\newlength{\figrulesep} 
\setlength{\figrulesep}{0.5\textfloatsep} 

\newcommand{\topfigrule}{\vspace*{-1pt}%
\noindent{\color{cream}\rule[-\figrulesep]{\columnwidth}{1.5pt}} }

\newcommand{\botfigrule}{\vspace*{-2pt}%
\noindent{\color{cream}\rule[\figrulesep]{\columnwidth}{1.5pt}} }

\newcommand{\dblfigrule}{\vspace*{-1pt}%
\noindent{\color{cream}\rule[-\figrulesep]{\textwidth}{1.5pt}} }

\makeatother

\twocolumn[
  \begin{@twocolumnfalse}
{\includegraphics[height=30pt]{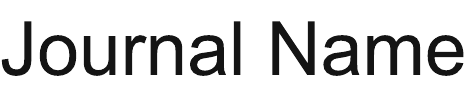}\hfill\raisebox{0pt}[0pt][0pt]{\includegraphics[height=55pt]{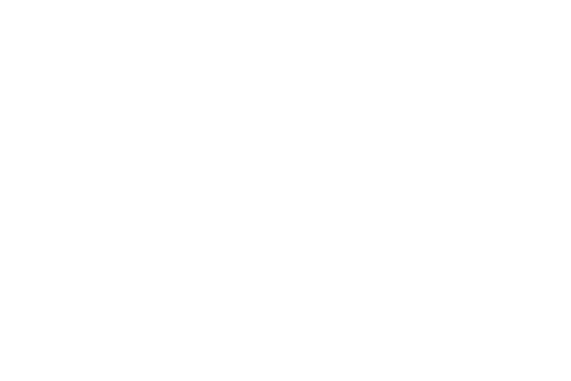}}\\[1ex]
\includegraphics[width=18.5cm]{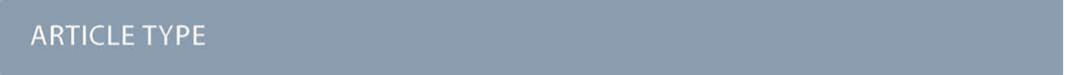}}\par
\vspace{1em}
\sffamily
\begin{tabular}{m{4.5cm} p{13.5cm} }

\includegraphics{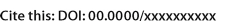} & \noindent\LARGE{\textbf{Tunable dynamics of flexible magnetic microcrosses: synchronous rotation, breathing and out-of-plane arm overtaking$^\dag$}} \\
\vspace{0.3cm} & \vspace{0.3cm} \\
 & \noindent\large{
 Joseph Tavacoli,\textit{$^{\ddag,a}$} 
 Andris P. Stikuts,\textit{$^{\ddag,b,c}$} 
 Mihir Dass,\textit{$^{a}$} 
Tim Liedl \textit{$^{a}$} and
 Pietro Tierno,\textit{$^{*,b, c}$}
 } \\
\includegraphics{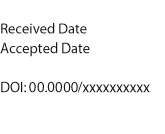} & \noindent\normalsize{We combine colloidal self-assembly and soft-lithography techniques to realize flexible magnetic microcrosses that can be manipulated via external, time dependent magnetic fields. 
The crosses are characterized by a central domain connected via four flexible arms. When subjected to an in-plane, rotating magnetic field, the crosses transit from a synchronous to an asynchronous spinning motion where their average rotation decreases with the driving frequency. In the asynchronous regime and at low field amplitudes, the crosses display a breathing mode, characterized by relative oscillations between the arms, while remaining localized in the two dimensional plane. In contrast, for high field amplitudes, we observe an arm overtaking regime where two opposite filaments surpass the remaining ones
forcing the cross to perform a 
three-dimensional gyroscopic-like rotation. Using slender body theory and balancing the effect of magnetic and elastic interactions, we recover the experimental findings and show that the overtaking regime occurs due to different arm magnetizations. Our engineered microscopic colloidal rotors characterized by multiple flexible filaments may find potential applications for precise lab-on-a-chip operations or as stirrers dispersed within microfluidic or biological channels. } \\
\end{tabular}

 \end{@twocolumnfalse} \vspace{0.6cm}
 ]

\renewcommand*\rmdefault{bch}\normalfont\upshape
\rmfamily
\section*{}
\vspace{-1cm}
\footnotetext{\textit{
$^{a}$~Faculty of Physics and Center for NanoScience
Ludwig-Maximilians-Universit\"at, Geschwister-Scholl-Platz 1, 
München 80539, Germany. E-mail: j.tavacoli@lmu.de
}}
\footnotetext{\textit{
$^{b}$~Departament de F\'{i}sica de la Mat\`{e}ria Condensada, Universitat de Barcelona, 08028, Spain.
E-mail: ptierno@ub.edu}}
\footnotetext{\textit{
$^{c}$~University of Barcelona Institute of Complex Systems (UBICS), 08028, Barcelona, Spain}}
\footnotetext{\dag~Electronic Supplementary Information (ESI) available: Five videos illustrating the dynamics of magnetic crosses in experiments and simulations. See DOI: 00.0000/00000000.}
\footnotetext{\ddag~J. T. and A. P. S. contributed equally to this work.}




\section{Introduction}
Micro and nanoscale devices able to convert internal or environmental energy into mechanical motion 
are becoming increasingly important for their direct applications
in diverse fields including microfluidics,~\cite{Terray2002,Ismagilov2003,Pamme2006,Sawetzki2008,Kavcic2009,Pedrero2015} microrheology,~\cite{Chevry2013,Chevry20132,Brasovs2015,Berret2016,Radiom2021} 
targeted drug delivery~\cite{Tierno2008,Medina2016,Srivastava2016,Zhang2010,Petit2012,Tierno2021,Junot2023}  and in-vivo healthcare.~\cite{Nelson2010,Bunea2020,Immihan2020,Soto2020} Additionally, they can exhibit startling collective behavior~\cite{Elgeti2015,Zottl2016,Bechinger2016,Illien2017} and their study offers insights in the unique dynamics and interactions at low Reynolds numbers, where inertial forces are negligible compared to viscous ones.~\cite{Happel1973,Purcell1977} Thus, the practical development and understanding of such devices is a crucial research focus. 

Due to their scale, these systems typically require specialized architectures for efficient operation, some of which have been mimicked from naturally occurring analogues. An example are flexible micron/nanoscale appendages.~\cite{Dreyfus2005,Maier2016,Magdanz2020,Pauer2021} These appendages can be employed in nature to direct the flow of extracellular fluid (cilia) and for motility purposes (flagella) where their flexibility is essential to create net flow/thrust over a stroke. Human-engineered versions have facilitated the realization of different adaptable micromotors, some of which demonstrate complex and guided mechanical function at the microscale and below.~\cite{Maier2016,Lak2024}  A diverse set of strategies have been employed for the operation of these  flexible microstructures, including the use of monochromatic light,~\cite{Palagi2016} acoustic waves,~\cite{Shaohua2023} electric~\cite{Yan2016,Nishiguchi2018} and magnetic fields.~\cite{Cebers2016,Jang2015,Liao2019} Since the latter represent a cheap, practical, programmable and biocompatible way to actuate over microscale matter, magnetic field manipulation techniques are particularly prevalent and promising for applications.

Here we add to this library a new set of magnetic microrotors that consist of four flexible appendages, in the form of an equal-angled cross. While flexible magnetic filaments have been previously engineered at the micronscale,~\cite{Furst1998,Furst1999,Goubault2003,Cohen2005,Dreyfus2005,Biswal2003,Gao2010,Li2010} this is the first case, to the authors’ knowledge, in which they have been co-joined at a four-way junction. By using an in-plane, circularly polarized, rotating magnetic field, we find a series of spinning configurations including a breathing mode with relative arm oscillations within a plane and an overtaking one, where the arms exchange their location during rotation forcing three dimensional motion (3D). We replicate these dynamical regimes via numerical simulations by considering the balance between magnetic, viscous and elastic forces acting on the four flexible arms. Moreover, we also find numerically that a sharp dynamical transition bridges the breathing and overtaking modes, when the  rotational dynamics of the microcrosses pass from being in-plane to a 3D gyroscopic-like motion. In short, the flexible nature of our crosses permits a broad range of programmable and functional motions, and a conformational flexibility which is completely absent in rigid analogues. 

\begin{figure}[t]
\centering
\includegraphics[width=\columnwidth,keepaspectratio]{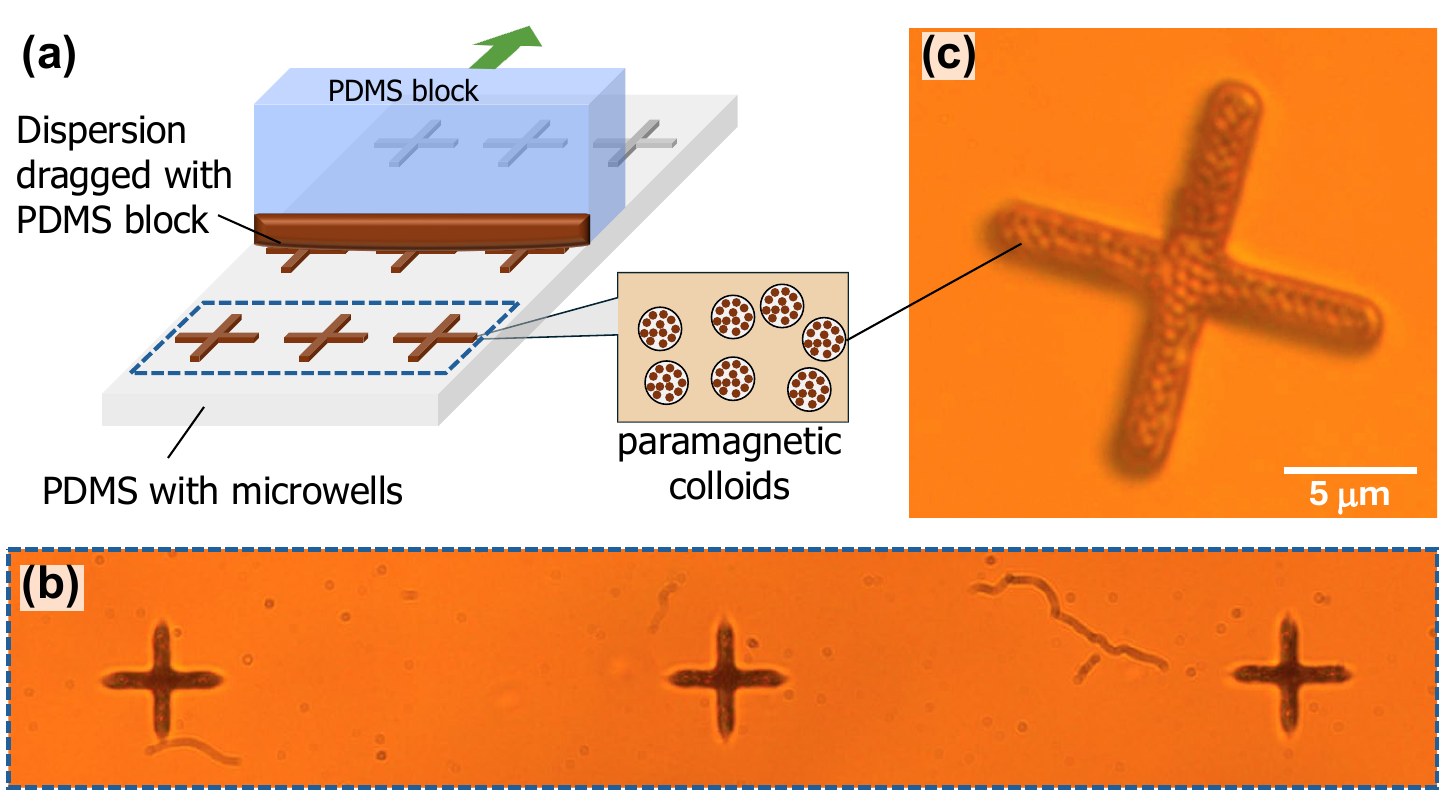}
\caption{(a) Schematic showing the fabrication method to realize the flexible magnetic microcrosses, combining soft lithography and colloidal self-assembly. (b) Microscope image showing three separate microcrosses within the PDMS block after being filled by the suspension of paramagnetic colloids ($1 \, \rm{\mu m}$ diameter).
(c) Image of one microcross with  
the embedded paramagnetic particles visible under an optical microscope.}
\label{figure1}
\end{figure}

\section{Experimental methods}
Our paramagnetic microcrosses are realized by embedding paramagnetic colloidal particles within lithographically designed structures. The process is schematically summarized in Fig.~\ref{figure1}(a), and it consists of the following steps. First, we use soft lithography to fabricate a rectangular polydimethylsiloxane (PDMS) mold characterized by separated holes having the shape of microcrosses. After that, we deposit on top of 
this substrate a $\sim 20 \rm{\mu L}$ droplet composed of a water suspension of paramagnetic colloids  (Dynabeads Myone, $1 \rm{\mu m}$ diameter) containing $0.1\%$ of Sodium Dodecyl Sulfate (SDS).
To fill the PDMS holes with this suspension, we slowly slide the colloidal suspension  using a block of PDMS, as shown in the schematic in Fig.~\ref{figure1}(a).
During sliding, we find that the particles are trapped within the 
microwells filling the square holes.
After evaporation of water, the particles strongly self-assemble within the mold due to attractive Van der Waals (VdW) interactions, Fig.~\ref{figure1}(b). 
Compared to previous procedures~\cite{Tavacoli2013,Junot2022},
here we did not use any monomer and cross-linking process, 
and found that VdW forces were able to stick irreversibly the particles within the molds.
The final product
consists of a microcross with four equal arms having a lateral size of $15\times 15 \rm{\mu m}^2$ with a thickness of $2\rm{\mu m}$, and completely filled with paramagnetic colloidal particles, Fig.~\ref{figure1}(c). 
In order to release the microcrosses, we cut a rectangular $4 \times 4$ mm segment from the filled PDMS mold containing many crosses. This segment was immersed in a $25 \, \rm{\mu l}$ solution of Polyvinylpyrrolidone (PVP, $10$k) in ethanol ($10$ wt\%) and centred on a glass slide until full evaporation of the solvent. The PDMS segment was then gently peeled away to leave the crosses embedded in a thin layer of PVP on the glass.
After this process, we place around the the crosses in PVP an additional $\sim 1$ mm deep frame of PDMS, of inner dimensions $5 \times 5$ mm, to hold the crosses after their release. The crosses were released from the polymer by gently pipetting on top of it a water solution containing a $0.1$ wt\% of SDS (aq). The system was finally sealed  by placing a glass cover slip on top of  the PDMS frame.

The experimental measurements were performed using an inverted optical microscope (Zeiss Axiovert, 100MHAL) equipped  with  two pairs of orthogonality aligned,  magnetic coils (inner diameter $35$ mm, outer diameter $57$ mm, height $15$ mm, $120$ turns with $0.85$ mm Cu wire, Express Transformers) having a surface-to-surface separation of $30$ mm. The spinning motion  of the magnetic crosses was induced via an external, rotating magnetic field, Eq.~\ref{field}, which was generated using a 200 W power amplifier (Kepco BOP20-10DL-802E) that was controlled by a wave function generator (Agilent, 372 33220A). The crosses where recorded using a  Thorlabs CMOS Camera (ThorlabsKiralux CS895CU) working at $20$ frames per second. 

\section{Results} 
\subsection{Flexibility of the magnetic crosses}
\label{sec:motion-regimes}

\begin{figure}[t]
\centering
\includegraphics[width=\columnwidth,keepaspectratio]{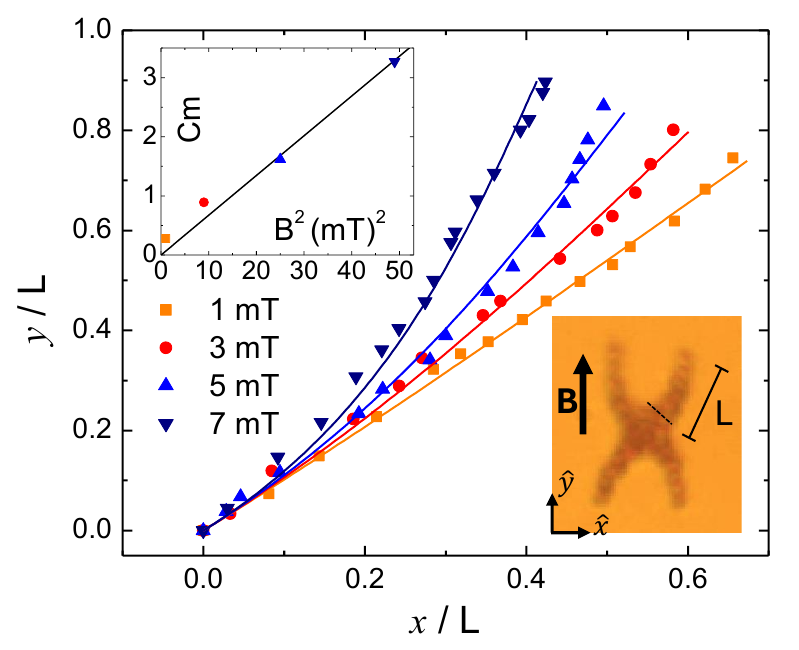}
\caption{Position of the magnetic filament in the ($x,y$) plane when subjected to a static magnetic field $B$ and averaged over different microscope images. Scattered symbols are experimental data,  continuous lines are non-linear regressions using Eq.34 in the Appendix. Bottom inset shows the image of one cross with a filament of length $L= 6.2\rm{\mu m}$ under a magnetic field $B=7$mT. The dashed line denotes the location of the starting point ($0,0$). Top inset shows the extracted magnetoelastic number $Cm$ versus field square $B^2$ with a regression, $Cm=\delta B^2$, where $\delta = 0.067 \pm 0.003 \, \rm{mT^{-2}}$.}
\label{figure2}
\end{figure}

Once dispersed in water, our magnetic crosses sediment close to the bottom of the experimental cell due to density mismatch and display  negligible thermal fluctuations. 
However, they do not stick to the surface, but can be oriented and deformed by external magnetic fields. 
We start by characterizing the shape deformation of the  arms composing a cross using a static field, which allows us to determine their flexibility. For a magnetic field applied along a defined direction, here 
$\bm{B}=B\hat{\bm{y}}$, inset in Fig.~\ref{figure2}, we observe a small bending of the arms towards the field direction along which the cross elongates. 
Using video-microscopy, we determine the shape changes of the arms in the $(x,y)$ plane for different field amplitudes and show them in Fig.~\ref{figure2}, normalized to the arm length $L= 6.2\rm{\mu m}$. Note that here these deformations are measured from the beginning to the end of the arm, not the center of the cross. 
The bending increases for high field amplitude and the cross reaches a lateral deformation of  $\sim 0.34 L=2.1 \rm{\mu m}$ under $7mT$ with respect to the undistorted case ($B=0$mT). 

The deformed shapes of the arms correspond to the equilibrium configuration under a static field, which can be calculated by  modeling the arms as elastic paramagnetic rods.\cite{Cebers2016} While more technical details will be given in Section 4 and in the Appendix, its physical behavior can be characterized in terms of the dimensionless magnetoelastic number:
\begin{equation}
Cm\equiv \frac{F_ML^2}{E_b} \, \, \, ,
\end{equation}
that provides the ratio between the magnetic to the elastic forces. Here
$F_M$ is the 
characteristic magnetic force acting on the cross-section of the arm, which can be estimated by considering the presence of paramagnetic colloidal particles embedded in it, resulting in the magnetic volume susceptibility $\chi$, $F_M=B^2\chi^2\pi a^2/\mu_0(\chi+2)$, with $a$ being the cross-section radius of the arm.
A detailed derivation of $F_M$ can be found in the Appendix.
The elastic contribution is characterized by the 
bending modulus $E_b$,
which describes the arm's stiffness or resistance to bending. Thus, by balancing the magnetic and the elastic forces,
one can calculate the arm's tangent angle $\theta$ as a function of its arc-length $l$
(see Eq.~\ref{angle-equation} in the Appendix) which can be used to fit the experimental data. From the non linear regressions, we 
extract the corresponding magnetoelastic numbers
that vary as $Cm\in[0.28,3.27]$ for $B\in [1,7]$ mT,
top inset in Fig.~\ref{figure2}.
Moreover, we find that $Cm$ grows linearly with $B^2$, 
as correctly predicted by our estimate of the filament magnetization. 

We estimate the magnetic volume susceptibility of the cross as $\chi=0.64\chi_b$, where $\chi_b=1.1$  \cite{Clime2007,Tierno20082} is the susceptibility of a single magnetic bead, which is multiplied by the random sphere packing fraction, $\phi=0.64$. 
We take the length of the arm as $L=6.2 \rm{\mu m}$, and its cross-section radius $a=1\rm{\mu m}$.
Using these values and the obtained slope in the inset of Fig.~\ref{figure2}, we can calculate the bending modulus of the arms as $E_b=2.6\cdot10^{-22} \rm{J\cdot m}$.
If we compare the obtained value with others present in the literature for flexible magnetic chains, we find some differences that emerge from the procedure used to obtain these chains and the chemical nature of the linkage between the particles. For example, our calculated bending modulus is four orders of magnitude larger  than that of
the flexible paramagnetic filaments produced by linking $\sim 600$ nm diameter particles with adsorbed polymer bridges, $E_b=7.8\cdot10^{-26} \rm{J\cdot m}$.~\cite{Goubault2003} This implies that our arms are much stiffer and they more effectively resist bending compared to thinner filaments. This is due to their larger lateral size that can accommodate several assembled particles, Fig.~\ref{figure1}(c). 
On the other hand, they are similar in flexibility to chains of paramagnetic particles of $1.1 \rm{\mu m}$ diameter  linked with 
poly-ethylene-glycol spacer molecules~\cite{Goubault2003}, where the bending modulus  was estimated to vary $E_b \sim 10^{-21} - 10^{-23}\rm{J\cdot m}$ depending on the used cross-linker.

\subsection{Magnetic manipulation: dynamical regimes}

\begin{figure}[t]
\centering
\includegraphics[width=\columnwidth,keepaspectratio]{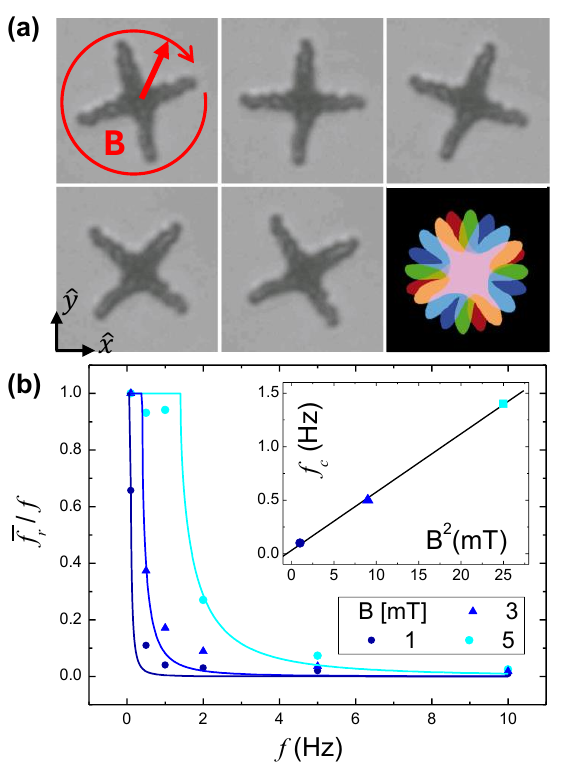}
\caption{(a) Sequence of optical microscopy images showing a microcross rotating in the $(\hat{\bm{x}},\hat{\bm{y}})$ plane due to a circularly polarized, rotating magnetic field with amplitude $B=3$mT and frequency $f= 10$ Hz. The sequence proceeds left to right then down, and each image is $0.3$ seconds apart.  (b) Normalized mean rotational frequency of the microcross, $\bar{f}_r/f$  
as a function of the driving frequency $f$. Scattered symbols are experimental data, continuous lines are non linear regressions following Eq.~\ref{adler} in the text. Inset shows the determined critical frequency $f_c$ versus square of the field amplitude $B^2$ with a linear regression.}
\label{figure3}
\end{figure}

Having characterized the crosses' flexibility, we proceed to investigate their dynamic behavior under 
a time dependent magnetic field. In particular,  
we drive these crosses using 
an in-plane, circularly polarized rotating magnetic field,
\begin{equation}
\bm{B}\equiv B [\cos{(2\pi ft)}, - \sin{(2\pi ft)}] \, \, \, ,
\label{field}
\end{equation}
with $B$ being the field amplitude and $f$ the driving frequency,
which we vary as $B\in [1,5] $ mT, $f\in [0.1,10]$ Hz. 
As shown in the sequence of images in Fig.~\ref{figure3}(a),
a microcross tries to follow the field dynamics and displays rotational motion around its center.
%
\begin{figure*}[t]
\centering
\includegraphics[width=\textwidth,keepaspectratio]{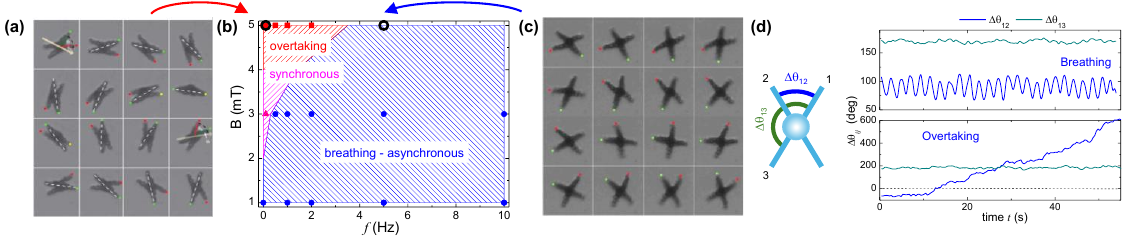}
\caption{(a,c) Sequence of optical microscope images showing the arm overtaking (a, time interval between images $\Delta t = 0.75$s) and the breathing (c, $\Delta t = 0.05$s) regimes. Two nearest arms of the crosses are highlighted by a red and a green disk. (b) Diagram in the $(f,B)$ plane showing the regions where overtaking (red) and breathing (blue) regimes are observed. 
In the first one, the cross rotation is synchronous (pink-red region), while the second one occurs when the crosses rotate asynchronously with the field. The boundaries of the diagram are guides to eye to distinguish between the two regimes.
The corresponding locations of the sequence of images in (a) and (c) are shown by the open black circles in the diagram. (d) Left: schematic showing a cross with four arms and the relative angles $\Delta \theta_{12}$ and $\Delta \theta_{13}$ between nearest and opposite filaments. Right: time evolution of the angles $\Delta \theta_{12}$ (blue) and $\Delta \theta_{13}$ (green) for the breathing (top) and overtaking (bottom) regimes 
obtained for rotating field with amplitude $B=5$mT and frequencies $f=5$Hz (breathing) and $f=0.1$Hz (overtaking).}
\label{figure4}
\end{figure*}
Thus, we measure the average rotational frequency $\bar{f}_r$
as a function of the field driving one, $f$. Similar to previous works on rotating magnetic colloids~\cite{McNaughton2006,Tierno2009,Coughlan2016} and ferrofluids holes~\cite{Helgesen1990} or droplets,~\cite{barcis1997} we observe two distinct regimes of motion, which result from the balance between magnetic and viscous torque acting over the crosses. 
\begin{figure*}[t]
\centering
\includegraphics[width=\textwidth,keepaspectratio]{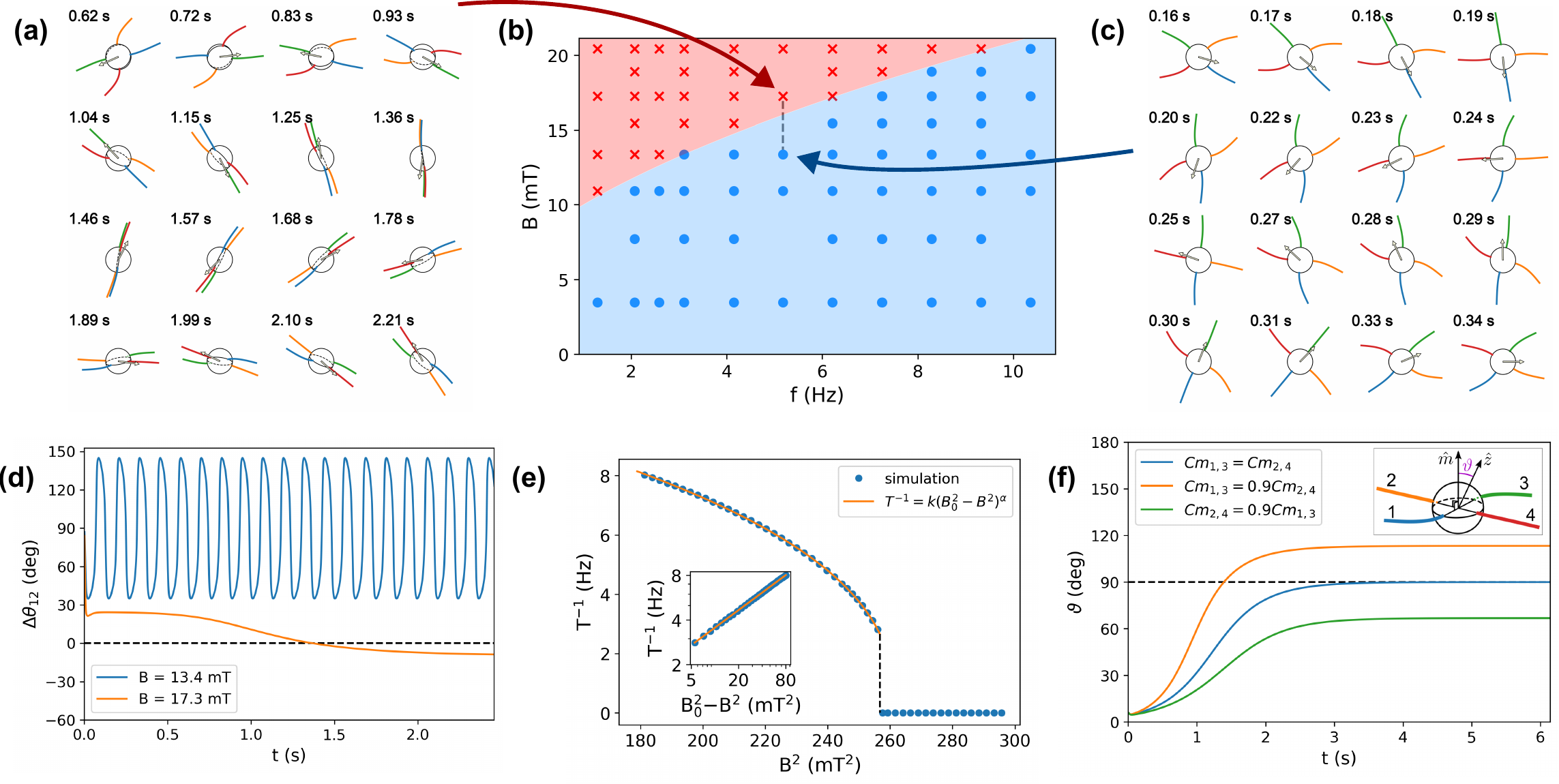}
\caption{Simulation results: (a,c) Snapshots from numerical simulations showing the relaxation to synchronous out-of-plane ($B=17.3$ mT, $f=5.2$ Hz) (a) and asynchronous in-plane breathing ($B=13.4$mT , $f=5.2$ Hz) (c) regimes. The white arrows in the crosses indicate the magnetic field direction, and the time is written on each snapshot. The blue and green arms have a $10\%$ smaller $Cm$ than the red and yellow arms. See also Movie S3 for (c) and Movie S4 for (a) in the Supporting Information. 
(b) Diagram in the ($B,f$) plane showing the regions where the synchronous out-of-plane motion (pink) and the asynchronous in-plane breathing (light blue) is observed. The red crosses and the blue disks show simulation results. The dark red and dark blue arrows show which points correspond to the (a) and (c) plots. 
(d) Time evolution of the angle $\Delta\theta_{12}$ between two adjacent arms as seen from the direction of the field rotation for the two different regimes. The field rotates with $f=5.2$ Hz. The dashed line shows where this angle is $0$, and thus where the cross tilts as illustrated in (a). In both regimes the angle $\Delta\theta_{13}=180^{\circ}$ is constant up to machine precision, thus it is not shown. 
(e) Inverse period $T^{-1}$ of the oscillation of $\Delta \theta_{12}$ versus $B^2$ measured along the transition path given by the dashed line in graph (b). Scattered disks are simulation results while continuous line is a non linear regression using power law with exponent $\alpha= 0.39$. The inset shows the log-log plot of $T^{-1}$ versus $B_0^2-B^2$, where $B_0^2 = 261.5\rm{ mT^2} $. 
(f) The polar angle $\vartheta$ of the cross as a function of time for the synchronous out-of-plane regime ($B=17.3\rm{ mT}, f=5.2 $Hz). Three different lines show the cases when all four arms have the same $Cm$ (blue), one opposite pair of arms has a $10\%$ smaller $Cm$, and the less magnetic pair is initially aligned with the field (orange, it is also illustrated in (a)), when one opposite pair of arms has a $10\%$ smaller $Cm$, but the more magnetic pair is now initially aligned with the field (green).
The dashed line shows $\vartheta=90^\circ$, where the cross is aligned perpendicular to the field rotation plane. See also Movie S5 in the Supporting Information.}
\label{figure5}
\end{figure*}
These regimes are
separated by a critical frequency $f_c$. For $f<f_c$ synchronous 
rotation occurs where the microcross rotates with field frequency $\bar{f}_r=f$
and its magnetic moment follows the applied field with a constant phase-lag angle. 
Due to the relatively high viscous torque acting over the four flexible arms, this effect is observed only at large amplitudes, $B\geq 3mT$ and at  low frequencies, $f<2$Hz, Fig.~\ref{figure3}(b).
In contrast, for  $f>f_c$ the crosses enter an  asynchronous regime where the viscous torque exceeds the magnetic one and the phase lag angle between the net magnetic moment and the rotating field is not constant but changes with time. This effect manifests as a rocking-like rotation superimposed on a continuous one. In this regime, the spinning frequency decreases and it can be described by the deterministic Adler equation:~\cite{Adler1946}
\begin{equation}
 \frac{\bar{f_r}}{f}=\begin{cases}
    1, & \text{if $f<f_c$}\\
    1-\sqrt{1-(f_c/f)^2}, & \text{if $f>f_c$} \, \, \, .
  \end{cases}
\label{adler}
\end{equation} 
Thus, by fitting the experimental data with Eq.~\ref{adler}, we determine the critical frequency
and show in the inset in Fig~\ref{figure3}(a) that it increases with $B^2$, a behavior similar to that observed for paramagnetic stiff ellipsoids under similar field modulations.~\cite{Tierno2009,Tierno2011}

Further, we find that the flexibility 
of the cross's arms introduces two different conformational adjustments, illustrated in Figs.~\ref{figure4}(a),(c), that are impossible for simple spherical or elongated particles. 
As shown in the diagram in Fig.~\ref{figure4}(b), we observe that in the asynchronous regime 
the crosses rotate as a whole at $\bar{f_r}<f_c$,
and the arms display relative  oscillatory displacements between them  
superimposed to the overall, rotational motion. We can characterize this breathing mode by measuring the time evolution of 
the relative angle between two arms, $\Delta \theta_{ij} =  \theta_i-\theta_j $ with $i\neq j$ as seen from above.
As shown in Fig.~\ref{figure4}(d), 
while the angle between opposite arms, $\Delta \theta_{13}$, remains almost constant and close to $180^{\circ}$,  the periodic oscillations of nearest tips, described by $\Delta \theta_{12}$, are also centred around $90^{\circ}$ but present a large amplitude of nearly $50^{\circ}$.
In this regime, the displacements of the crosses occur mainly within the confined plane, see Supplementary video 1. This effect contrasts with the second regime observed at high field amplitudes and low frequencies, Fig.~\ref{figure4}(b). Now the crosses perform a three dimensional 
spinning-like motion, where two opposite arms  overtake the remaining ones forcing the flipping of the magnetic structure, to keep the cross synchronous with the rotating field, see also Supplementary video 2. 
This switching between arms' locations emerges in $\Delta \theta_{12}$ as a continuous change in its 
value, which starts from $\sim -90^{\circ}$ and then it changes sign crossing the value of $\Delta \theta_{12}=0^{\circ}$ where both arms overlap, 
bottom of Fig.~\ref{figure4}(d).
Both dynamical regimes result from the presence of elasticity and multiple, interacting  elements that try to follow the applied magnetic torque. 
Clearly for a perfectly symmetric system, and in absence of any source of noise or tilt,
it would be difficult to observe any out of plane movement such as the 
arm overtaking, and the crosses would rather follow synchronously the rotating field. 
However, in this regime  the magnetic crosses flip in 3D to allow the overlapping and surpassing between pair of opposite arms  as viewed from above. This effect points towards the presence of a slight unsymmetrical filling of the arms with paramagnetic colloids during to the fabrication process. To confirm this effect,
and to reproduce the different regimes of motion, 
we next 
put forward a theoretical model 
incorporating magnetic, elastic and viscous forces and that takes into account also the possibility of this difference in the magnetization between opposite arms.

\section{Theoretical model}
\label{sec:theor}
We consider a magnetic cross as composed of four paramagnetic elastic filaments with a cross-section radius $a$, length $L$, bending modulus $E_b$, and volume magnetic susceptibility $\chi$. The filaments  are attached to a central structure, which for simplicity we model as a sphere with radius $A$.
The cross is placed inside a fluid of viscosity $\eta$ and subjected to a uniform rotating magnetic field, Eq.~\ref{field}. 
We model the coupling of the arms to the central sphere in a similar way to a previous work on a Quincke roller with an attached filament.~\cite{Zhu_Stone_2020} 
The central sphere rotates with the angular velocity $\bm\Omega$ and translates with the velocity $\bm U$ according to the force and torque balance equations:
\begin{equation}
\label{eq:sphere_omega}
    \sum_{i=1}^4 \bm T^{f_i\rightarrow s} - 8\pi\eta A^3 \bm \Omega = \bm 0,
\end{equation}
\begin{equation}
\label{eq:sphere_v}
    \sum_{i=1}^4 \bm F^{f_i\rightarrow s} - 6\pi\eta A \bm U = \bm 0,
\end{equation}
where $\bm F^{f_i\rightarrow s}$ and $\bm T^{f_i\rightarrow s}$ are, respectively,  the force and the torque exerted by the filament $i$ on the sphere. Under a uniform magnetic field a sphere experiences no force or torque as it should be for a paramagnetic colloid.

We assume that the magnetic filaments follow the Kirchoff model of an elastic rod, where magnetic interactions are included.~\cite{Cebers2016}
We describe the shape of the arms using their centerlines $\bm{r}(l)$ parametrized by the arc-length $l\in [0,L]$, where $l=0$ is located at the surface of the sphere.
The force acting on the cross-section of each arm is then:
\begin{equation}
    \bm{F}=-E_b \bm r_{lll}-F_M (\bm r_l\cdot \bm b)\bm b+\Lambda \bm r_l,
\label{eq:fil_force}
\end{equation}
where the subscript $l$ indicates the derivative with respect to the arc-length.
Here $F_M$ is the magnetic parameter defined in section \ref{sec:motion-regimes} and derived in the Appendix, $\bm b=\bm B/B_0$ is the unit vector along the magnetic field direction and 
$\Lambda(l)$ is the tension force ensuring the inextensibility of the arms.
The torque acting on each cross-section is given by: 
\begin{equation}
    \bm T = E_b \bm r_l \times \bm r_{ll} \, \, \, .
\end{equation}
We employ the resistive force theory to obtain the velocity of the arms:~\cite{lauga_fluid_2020}
\begin{equation}
    \bm v= \frac{(\bm F_l\cdot \bm r_l)\bm r_l}{\zeta_\parallel} + \frac{\bm F_l - (\bm F_l\cdot \bm r_l)\bm r_l}{\zeta_\perp},
\label{eq:fil_velocity}
\end{equation}
where $ \zeta_\parallel=2\pi\eta/(\ln{\left( L/a \right)} - 1/2)$  and $\zeta_\perp=4\pi\eta/(\ln{\left( L/a \right)} + 1/2)$ are the drag coefficients along the tangential and the perpendicular directions of the arms.~\cite{Blake2010}
Further we use dimensionless parameters given by the characteristic length scale $L=6.2\, \rm{\mu m}$, time scale $\zeta_\perp L^4/E_b = 0.03$ s and force scale $E_b/L^2=6.8$ pN.
More details on the model like the equation for the tension force $\Lambda(l)$, the expressions for the forces $\bm F^{f_i\rightarrow s}$ and torques $\bm T^{f_i\rightarrow s}$ on the sphere, the boundary conditions and the numerical simulation parameters are given in the Appendix.

\section{Discussions}

From the numerical simulations, details in appendix~\ref{simul}, we observe similar dynamical regimes as reported in the experiments. 
In particular, in physical units  we obtain a similar state diagram shown in  Fig.~\ref{figure5}(b). For low amplitude of the rotating  field and at large driving frequencies, we 
retrieve the breathing-like oscillations of the arms,
which occur within the blue region of the diagram.
Fig.~\ref{figure5}(c) displays a sequence of images of the simulated magnetic cross which shows the periodic widening and closing of nearest arms 
giving rise to the breathing motion, see also Movie S3 in the Supporting Information. 
Note that even if we initially slightly tilt the cross in the simulations, the magnetic structure in this regime relaxes back to the plane of rotation of the magnetic field and stays confined in two dimensions.
In contrast, for strong field amplitudes and at low frequencies, as denoted by the red region in Fig.~\ref{figure5}(b), we find that in our simulations 
the crosses extend and follow synchronously the rotating field. However, as time proceeds the in-plane motion becomes unstable and the cross tilts until it reaches a stable out-of-plane configuration, where it still rotates synchronously with the driving field.
This situation is illustrated in the sequence of images  in Fig.~\ref{figure5}(a) and also Movie S4 in the Supporting Information.
The corresponding relative angle, $\Delta \theta_{12}$, shown in Fig.~\ref{figure5}(d),
shows a similar trend to the experimental one, which allows us to distinguish 
both regimes and the transition from the in plane (2D) motion to the overtaking (3D) one. While in the breathing mode $\Delta \theta_{12}$ displays periodic oscillations (blue line), it switches sign in the overtaking, passing from $90^{\circ}$ to a negative value due to the exchange of position of the two arms. In contrast, we find that in both regimes $\Delta \theta_{13}=180^{\circ}$ for all simulation time steps.

Another advantage of the numerical simulations is that they allow a fine characterization of the transition from one dynamical regime to the other without any source of external noise or disorder within the system.
In particular, in Fig.~\ref{figure5}(e)
we perform a series of simulations by keeping the frequency constant at $f=5.2$Hz, and increasing the amplitude from $B=13.4$mT to $B=17.3$mT, as illustrated by the dashed line in Fig.~\ref{figure5}(b).
We monitor the frequency of the oscillations of $\Delta\theta_{12}$
in the breathing mode, given by the inverse of its period $T^{-1}$, that we use as order parameter.  We find that $T^{-1}$ decreases 
as the field increases starting from twice the 
field frequency $f$ for $B$ values close to 0, until reaching $0.51f$ at a transition field of  $B_0=16.0$mT. For this field value, the breathing oscillation frequency, drops to zero sharply and then it remains so for all other explored amplitudes. Close to the the transition field, the breathing  frequency  follows a power law behavior, as shown in the inset of Fig.~\ref{figure5}(e), with an exponent  $\alpha = 0.39$ which is close to the classical mean field exponent, $\beta = 1/2$~\cite{Kuramoto1984}.

Apart from the reduction of the breathing frequency, another feature of the transition from breathing to arm overtaking is the change from a confined 2D rotation to a 3D one.   
To characterize this effect, we define the polar angle $\vartheta$ of the cross as the angle that the axis (denoted by $\hat m$) perpendicular to the lines connecting the base points of the arms makes with the field rotation direction $\hat z$,
see the small inset at the top right in Fig.~\ref{figure5}(f). For an in-plane rotating cross, $\vartheta = 0^{\circ}$.
Since in simulation we can change the 
magnetic content of the arms, in Fig.~\ref{figure5}(f) we explore the
effect of varying the magnetoelastic number $Cm$
between pairs of opposite arms and how it affects the cross dynamics in terms of the time evolution of the polar angle $\vartheta$.
Note that, when all of the arms of the cross have the same magnetoelastic number $Cm$, the out-of-plane configuration results in $\vartheta=90^{\circ}$, which 
corresponds to a cross rotating with the arm pairs exactly one on top of the other.
If one opposite pair of arms has a larger $Cm$ than the other, the cross tilts to $\vartheta\neq90^{\circ}$, since the pair with the larger $Cm$ tends to align more with the field direction.
If the initial configuration is such that the field is more aligned with the less magnetic arms, the cross flips when reaching $\vartheta>90^{\circ}$. By doing this, the more magnetic pair of arms overtakes the less magnetic one and it aligns closer with the field direction.
An example of this motion is illustrated in Fig.~\ref{figure5}(a).
If initially the field is more aligned with the more magnetic arms, no such overtaking occurs and the cross relaxes to $\vartheta<90^{\circ}$.
These behaviors can be also observed in the sequence of video clips in  Movie S5 in the Supplementary Information.
Thus, we have shown numerically that the overtaking regime results from the different  magnetoelastic numbers of the two pairs of arms, which may result from 
the fabrication process. In particular, we believe that 
the dragging direction of the drop containing the magnetic particles on the PDMS mold (Fig.1~\ref{figure1}(a)) is crucial to the arm magnetoelastic properties. 
We found that the arms orientated perpendicular to the dragging direction were slightly more filled with particles due to capillarity.
In principle one can change the drop dragging direction in order to tune the amount of magnetic content in the arms and 
change the overtaking regime correspondingly. Investigating how the filling direction directly affects the crosses dynamics and their transition from in-plane to 3D motion may be an interesting venue for future research.

Finally we comment on the difference between our experimental results and the results from numerical simulations. In the experiment we see that for strong fields and low rotation frequencies, the crosses display 3D, out-of-plane motion. 
While from simulations we also find that the combined action of elastic, magnetic and viscous torques are sufficient to tilt the cross out of the rotation plane, in the experiment this motion typically does not reach an out-of-plane steady state. In contrast, the cross displays periodic tumbling (Movie S2 in Supplementary Information), or it remains confined to the rotation plane while moving synchronously with the field.
We hypothesize that the periodic tumbling results from the increase of the friction closer to the glass surface of the sample. Indeed, the slow rotation of the field makes the cross stand up, but the increased drag on the lower parts makes it fall down again.
In contrast, the cases where the cross rotates synchronously on the surface is due to the gravity force, which was also not included in the simulations.
In addition to that, while the trends of the two regimes in the diagrams in Fig.\ref{figure4}(b) and \ref{figure5}(b) are similar, the values of the corresponding physical parameters are different. Indeed, in experiments we use field amplitudes $B \in [1,5]$mT and frequency vales $f\in[0.1,10]$Hz which are smaller than the simulation ones ($B \in[3.5,20.4]$mT, $f\in[1.0,10.4]$Hz). Even if we employ flexibility value and magnetoelastic number extracted from the experimental data, Fig.\ref{figure2}, still the discrepancy arises due to the different approximations used in our model.
In simulations, the magnetic field necessary to keep the microcross in synchronous rotation is approximately twice larger than that used in the experiments. 
We can attribute this to an overestimate of the drag force in the simulation possibly due to the quite large central spherical piece compared to the thickness of the arms. 
Additionally, we did not consider the presence of the close surface and the corresponding change in the drag coefficient when the crosses rotate close to it in the breathing or overtaking regimes. Also, we neglect gravity and any types of imperfection/disorder present in our system. Despite these approximations, we still capture qualitatively well the different dynamical regimes emerging from our magnetically driven microcrosses.

\section{Conclusion}
We have experimentally realized microscale magnetic crosses
using a combination of soft lithography and particle self assembly. When driven by an in-plane rotating field, these microstirrers show a sequence of dynamical regimes which emerge as a consequence of the balance between magnetic torque, elasticity and viscous dissipation.
In particular, apart from displaying the transition from synchronous to asynchronous dynamics, we observe a breathing mode characterized by relative arm oscillations and an overtaking one where the arms can overpass each other due to their different magnetizations. 
We have modeled the crosses by considering four flexible, paramagnetic arms attached to a rigid spherical body, and have been able to recover all the observed regimes taking into account the different approximations used, including the absence of the close substrate.

The controlled, rotational motion of flexible magnetic filaments in water has been the subject of 
intense research so far, due to its direct application in microstirring and microfluidics.
Here we have realized a composite structure where the 
filaments are not  able to spin individually, but they are part of a larger body and collectively take part in its rotational motion. 
Thus, we expect that our crosses could provide additional functionality due to their extended surface and 3D motion within 
microfluidic channels when used as field controllable inclusions or 
in microrheological studies to investigate the viscoelastic properties of 
complex fluids. 
In addition to that, an exciting extension of our work may be to investigate the dynamics of collective ensembles of magnetically driven microcrosses and the corresponding emerging states. The dynamic assembly of in-plane magnetic spinners has been 
investigated in the past by different research groups reporting novel phases and emerging stationary states~\cite{Grzybowski2000,Grzybowski2002,Jozef2008,Yan2015,Kokot2017,Soni2019,Petrichenko2020,Kokot2020,Raul2021}. 
Introducing elasticity within the interacting units may provide an additional 
degree of freedom to play with, leading to a rich state diagram 
and non-trivial collective rotational dynamics. Thus, plans to explore many interacting crosses by varying both the amount of elasticity and system density  
are on the way.

\section{Appendices}

\subsection{Magnetic energy}

The energy associated with a paramagnetic body in an external field is: \cite{stratton2007electromagnetic}
\begin{equation}
U_M=-\frac{1}{2}\int_V \vect M \cdot \vect B_{ext} dV,
\end{equation}
where $\vect M$ is the induced magnetization of the body, $\vect B_{ext}$ is the applied magnetic field, and the integral is taken over the volume of the paramagnetic body.
Now let us consider the associated energy of a short segment of the paramagnetic filament oriented along the tangent $\vect r_l$ direction. 
For a short segment characterized by a small cross section, we can write its magnetization similar to that of a slender ellipsoid
\begin{equation}
    \vect M_{\parallel} = \frac{\chi}{1+\chi N_{\parallel}}\frac{\vect B_{ext\parallel}}{\mu_0}, \quad \vect M_{\perp} = \frac{\chi}{1+\chi N_{\perp}}\frac{\vect B_{ext\perp}}{\mu_0},
\end{equation}
where $\parallel$ denotes the direction along $\vect r_{l}$ and $\perp$ - the direction perpendicular to it. $\chi$ is the susceptibility and $N$ is the demagnetizing factor.\cite{stoner1945xcvii}. For a slender ellipsoid of revolution, we have $N_\parallel=0$ and $N_\perp=1/2$. 
This gives for the magnetic energy:
\begin{equation}
    \Delta U_M=-\frac{1}{2\mu_0}\int_{\Delta V} \left( \chi  B_{ext\parallel}^2+ \frac{\chi}{1+\frac{\chi}{2}}B_{ext\perp}^2 \right) dV,
\end{equation}
where the integral is over the volume $\Delta V$ of the small segment.
Using $B^2=B_\parallel^2+B_\perp^2$ we can transform it to 
\begin{equation}
    \Delta U_M=-\frac{1}{2\mu_0}\int_{\Delta V} \left(  
    \frac{\chi^2}{\chi+2}B_{ext\parallel}^2
    \right) dV + const,
\end{equation}
where the $const$ term contains the part of magnetic energy proportional to $B_{ext}^2$ that does not depend on the segment's orientation, and thus whose variation will not correspond to any torques and forces.
Therefore, in the following we omit it.
Assuming constant properties over the cross-section of the filament $\pi a^2$, we integrate over it, and writing $B_{ext\parallel}^2=(\vect B_{ext}\cdot \vect r_l)^2$, we get the linear magnetic energy density 
\begin{equation}
    \frac{d U_M}{dl} = -\frac{\pi a^2}{2 \mu_0} \frac{\chi^2}{\chi+2} (\vect B_{ext}\cdot \vect r_l)^2
\end{equation}
The magnetic energy for the whole filament then becomes 
\begin{equation}
    U_M = -\frac{F_M}{2}\int_0^L (\vect b\cdot \vect r_l)^2dl,
\label{eq:fil_energy}
\end{equation}
where $\vect b= \vect B_{ext}/B_{ext}$, and we combine the many parameters into the characteristic magnetic force on the cross-section:
\begin{equation}
    F_M=\frac{\pi a^2\chi^2B_{ext}^2}{\mu_0(\chi+2)} \, \, \, .
\end{equation}
Note that, changing from CGS to SI units, this expression corresponds to that used in Ref. \cite{Cebers2016}.

\subsection{Magnetic force on the cross-section}

Making a small variation in $\vect r\rightarrow\vect r+\delta \vect r$, under the constraint that the filament is inextensible ($\delta (dl) = 0$), we get:
\begin{equation}
    \delta U_M = -F_M\int(\vect b\cdot \vect r_l)(\vect b\cdot(\delta\vect r)_l)dl \, \, \, .
\end{equation}
We can integrate by parts to obtain
\begin{equation}
    \delta U_M = [-F_M(\vect b\cdot \vect r_l)\vect b\cdot \delta \vect r]+F_M\int(\vect b\cdot \vect r_{ll})\vect b\cdot \delta\vect rdl,
\label{eq:energy_variation}
\end{equation}
where the square brackets indicate the evaluation at the ends of the filament with the appropriate signs.
The change in energy corresponds to the work done on the filament \cite{Cebers2016, powers2010dynamics}
\begin{equation}
    \delta U = [\vect F\cdot \delta\vect r] + [\vect T\cdot\delta\vect\phi] - \int \vect F_l\cdot \delta \vect r dl,
\label{eq:work_on_filament}
\end{equation}
where $\delta\vect\phi$ is the rotation angle of the tangent vector.
Comparing Eqs. \eqref{eq:energy_variation} and \eqref{eq:work_on_filament}, we can read out the magnetic force
$ -F_M(\vect b\cdot \vect r_l)\vect b$, which we have in eq. (6) in the main manuscript.

\subsection{Details of the theoretical model.}
In the following equations we use dimensionless variables based on the scales introduced in Section \ref{sec:theor}. 
The equation for the tension force $\Lambda(l)$ is obtained from the local inextensibility constraint $\bm r_l \cdot \bm r_l=1$, whose time derivative gives $\bm v_l \cdot \bm r_l=0$. 
Plugging in equations \eqref{eq:fil_force} and \ref{eq:fil_velocity}, we get
\begin{equation}
\begin{split}
    -\Lambda \bm r_{ll}\cdot \bm r_{ll}&+\frac{\zeta_\perp}{\zeta_\parallel} \Lambda_{ll} = \\
    &3\left(\frac{\zeta_\perp}{\zeta_\parallel}-1 \right)\left( \bm r_{llll} \cdot \bm r_{ll} +\bm r_{lll}\cdot\bm r_{lll}  \right)+\\
    &4 \bm r_{llll} \cdot \bm r_{ll} + 3\bm r_{lll}\cdot\bm r_{lll} - \beta (1-\bm r_l\cdot \bm r_l),
\end{split}
\end{equation}
where the final term $-\beta (1-\bm r_l\cdot \bm r_l)$ is a added by hand to penalize numerical deviations from the inextensibility constraint~\cite{TORNBERG20048}.
In the simulations we use $\beta=10^5$.

We now explain in detail the boundary conditions used for the arms.
At the free end of each arm $l=1$, we have zero torque and zero force condition, which gives for the centerline: 
\begin{equation}
    \bm r_{ll}=0, \quad \bm r_{lll} = -Cm(\bm b\cdot\bm r_l)\bm b + Cm (\bm b\cdot\bm r_l)^2\bm r_l,
\end{equation}
and for the tension force: 
\begin{equation}
    \Lambda = Cm \left(\bm b \cdot \bm r_l\right)^2 \, \, \, . 
\end{equation}
Starting from the base at $l=0$,  the filament  is clamped in the sphere along a direction given by the unit vector $\bm e^i_p$ different for each filament $i$.
This gives the boundary conditions for the centerline
\begin{equation}
    \bm r = \bm r_s + \alpha\bm e_p^i, \quad  \bm r_l=\bm e_p^i,
\end{equation}
where $\bm r_s$ is the center of the sphere, and $\alpha=A/L$ is its dimensionless radius.
To obtain the boundary condition on the sphere $l=0$ for the tension force, we take a dot product of \eqref{eq:fil_velocity} with $\bm e_p^i$ to get: 
\begin{equation}
    \bm v\cdot\bm e_p^i =  3\frac{\zeta_\perp}{\zeta_\parallel}\bm r_{lll}\cdot \bm r_{ll} + \frac{\zeta_\perp}{\zeta_\parallel}\Lambda_l - \frac{\zeta_\perp}{\zeta_\parallel}Cm(\bm b\cdot\bm r_{ll})(\bm b\cdot\bm e_p^i).
\end{equation}
The center of the sphere $\bm r_s$ and the attachment points of the filaments move together with the sphere:
\begin{equation}
    \frac{d \bm r_s }{dt}=\bm U, \quad \frac{d \bm e^i_p}{dt} = \bm \Omega\times \bm e_p^i.
\end{equation}
The problem is finished by giving the expressions for the force and torque exerted by the filaments on the sphere 
\begin{equation}
    \bm F^{f_i\rightarrow s} = -\bm r_{lll}+\Lambda \bm e_p^i - Cm \left(\bm e_p^i\cdot \bm b \right)\bm b,
\end{equation}
\begin{equation}
    \bm T^{f_i\rightarrow s} = -\alpha \bm e_p^i \times \left( \bm r_{lll} + Cm \left(\bm e_p^i\cdot \bm b \right) \bm b \right) + \bm e_p^i \times \bm r_{ll},
\end{equation}
and these expressions are evaluated at the base of the filament, where $l=0$.

\subsection{Equilibrium configuration in a static magnetic field}
For static configurations the force acting on each cross-section of the arm is zero,
\begin{equation}
    \bm F = -\bm r_{lll} - Cm( \bm r_l \cdot  \bm b)\bm b + \Lambda
\bm r_l=\bm 0.
\end{equation}
We assume planar configuration of the arms along the direction of the magnetic field $\bm b=\{0,1,0\}$, therefore we write the tangent and normal vectors of the arm as $\bm t=\bm r_l = \{\cos \theta, \sin \theta, 0 \}$, and $\bm n = \{-\sin \theta, \cos\theta, 0\}$.
Then we have
\begin{equation}
    \theta_l^2 \bm t - \theta_{ll}\bm n - Cm \sin \theta \bm b + \Lambda \bm t=\bm0.
\label{eq:force_eqlib}
\end{equation}

To calculate the shape of the arms,
we project Eq. \eqref{eq:force_eqlib} on the $\bm n$ direction, and get: 
\begin{equation}
    \theta_{ll} + \frac{Cm}{2} \sin 2\theta = 0.
\end{equation}
We multiply by $\theta_l$ and integrate over $l$ to obtain
\begin{equation}
    \theta_l^2 - \frac{Cm}{2} \cos2\theta = c^2,
\label{eq:first_integral}
\end{equation}
where $c^2$ is a constant.
We can rearrange the equation to get the following
\begin{equation}
    dl = \frac{1}{\sqrt{c^2+\frac{Cm}{2}}} \frac{d\theta}{\sqrt{1-\frac{Cm}{c^2+\frac{Cm}{2}}\sin^2\theta}},
\end{equation}
which we integrate
\begin{equation}
    \int_0^l dl' = \frac{1}{\sqrt{c^2+\frac{Cm}{2}}} \int_{\theta_0}^{\theta}\frac{d\theta'}{\sqrt{1-\frac{Cm}{c^2+\frac{Cm}{2}}\sin^2\theta'}}.
\end{equation}
The result is expressed in terms of the elliptic integrals of the first kind $F(\theta,k)$
\begin{equation}
    l = \frac{k}{\sqrt{Cm}}\left[ F(\theta,k) - F(\theta_0,k) \right],
\label{arc-equation}
\end{equation}
where we have defined a new constant $k=\sqrt{Cm/(c^2+Cm/2)}$, and $\theta_0$ is the angle at the base of the filament, $l=0$.
The solution for $\theta(l)$ can be expressed using the Jacobi amplitude function $am$:
\begin{equation}
    \theta = am\left( \frac{\sqrt{Cm}l}{k}  + F(\theta_0,k), k\right).
\label{angle-equation}
\end{equation}
Similar expressions have been derived by minimizing the magnetoelastic energy in previous works for other systems~\cite{Shcherbakov2004,Erglis2008}.

What remains is to determine the value for $k$, which we get from the free end boundary condition $\theta_l(1)=0$.
Plugging it into the equation \eqref{eq:first_integral}, we get
\begin{equation}
    c^2=-\frac{Cm}{2} \cos(2\theta(1)) \, \, \, .
\end{equation}
We can rewrite it to obtain the equation that $k$ must satisfy
\begin{equation}
    k\sin\theta(1)=1.
\end{equation}
To obtain the tension force, we project the equation \eqref{eq:force_eqlib} onto the $\bm t$ direction as:
\begin{equation}
    \Lambda = Cm \sin^2\theta - \theta_l^2 \, \, \, .
\end{equation}
Plugging in the expression for $\theta$, we arrive at
\begin{equation}
\label{eq:Lambda-equation}
    \Lambda = \frac{Cm}{k^2}\left[ 1-2 dn\left(\frac{\sqrt{Cm}l}{k} + F(\theta_0,k)  ,k \right)^2 \right],
\end{equation}
where we used the Jacobi $dn$ function.

\begin{figure}[t]
\centering
\includegraphics[width=\columnwidth,keepaspectratio]{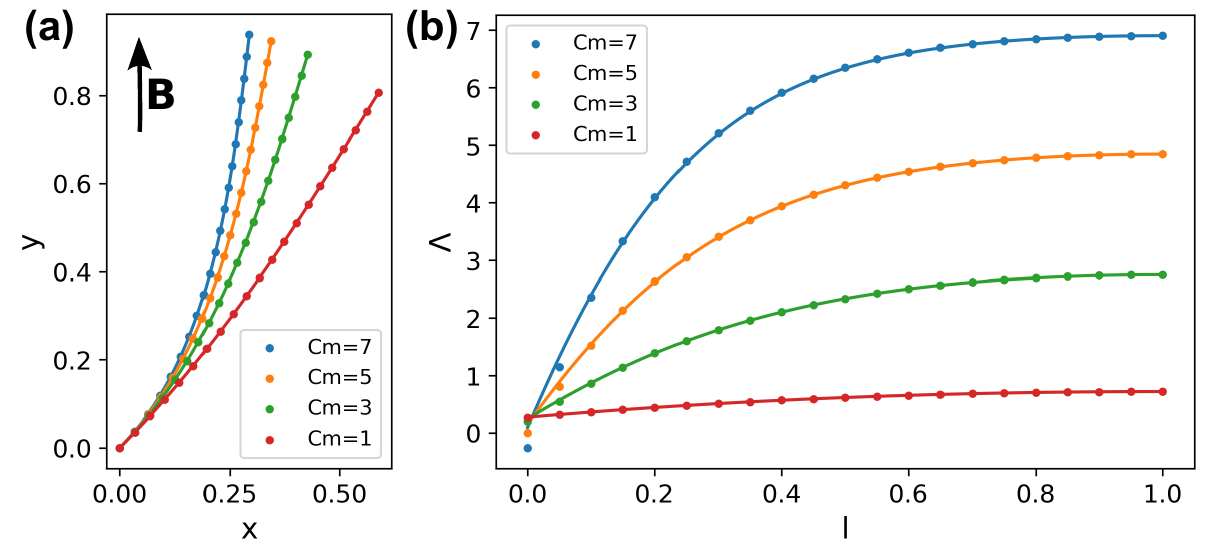}
\caption{(a) Equilibrium shape of a magnetic arm of a cross in a static vertical field for different magnetoelastic numbers $Cm$. The disks are results from numerical simulations and the lines are obtained from Eq.~\ref{angle-equation}. (b) The equilibrium tension force $\Lambda$ in an arm as a function of the arc-length $l$. The disks are simulation results and the lines are obtained from Eq.~\ref{eq:Lambda-equation}.  }
\label{figure6}
\end{figure}

\subsection{Details of the numerical simulations.}
\label{simul}

We numerically solve Eq.~\ref{eq:fil_velocity} to calculate the velocity of the arms  and  Eqs.~(\ref{eq:sphere_v},\ref{eq:sphere_omega})
for the velocity and angular velocity of the sphere. The arm positions are then numerically calculated using the backwards Euler method, and the simple Euler method for the sphere. Second order finite differences are used to calculate the derivatives with respect to $l$.
The constants used in the simulations are $a/L=0.1$, $
\alpha=A/L=0.4$, the arms are discretized into $N=21$ points and the time step is $\Delta t=3.3\cdot10^{-7}$. 
The initially straight arms are connected to the sphere every $90^\circ$ around its equator.
To break the in-plane symmetry, we orient the sphere such that one opposite pair of arms is tilted by $0.1 \rm{rad} \approx 5.7^{\circ}$ from the field rotation plane.
Finally, inspired by the experiment, we set the magnetoelastic number $Cm$ of one pair of arms to be $10\%$ smaller than the opposite one.
The simulation program was validated by comparing the calculated equilibrium configuration and the tension force of the arms in a static field to the analytic solutions Eqs.(\ref{angle-equation},\ref{eq:Lambda-equation}) resulting in an excellent agreement
between them as shown in Fig.~\ref{figure6}.

\section*{Data availability}
The data that support the findings of this study are available
from the corresponding authors upon reasonable request.

\section*{Conflicts of interest}
There are no conflicts to declare.

\section*{Acknowledgements}
This project has received funding from the European Research Council (ERC) under the European Union's Horizon 2020 research and innovation program (grant agreement no. 811234).
J. T. acknowledges the German Research Foundation (DFG) under project num. TA 1375/2-1.
P. T. acknowledges support from the 
Ministerio de Ciencia, Innovaci\'on y Universidades (grant no. PID2022-
137713NB-C21 AEI/FEDER-EU) and the Generalitat de
Catalunya (ICREA Acad\'emia).



\balance


\bibliography{Bibliography}
\bibliographystyle{rsc} 

\end{document}